# Effect of severe plastic deformation realized by rotary swaging on the mechanical properties and corrosion resistance of near-α-titanium alloy Ti-2.5Al-2.6Zr

V.N. Chuvil'deev[a], V.I. Kopylov[a,b], A.V. Nokhrin[a(*)], P.V. Tryaev[c], N.Yu. Tabachkova[d], M.K. Chegurov[a], N.A. Kozlova[a], A.S. Mikhaylov[c], A.V. Ershova[c], M.Yu. Grayznov[a], I.S. Shadrina[a], C.V. Likhnitskii[a]

[a] Lobachevsky State University of Nizhny Novgorod, 603950, Russia, Nizhny Novgorod, Gagarina ave., 23

[b] Physics and Technology Institute, National Academy of Sciences of Belarus, 220141, Belarus, Minsk, Kuprevich st., 10

[c] Afrikantov OKBM JSC, Russian Nuclear Corporation "ROSATOM", 603074, Russia, Nizhny Novgorod, Burnakovsky proezd, 15

[d] National University of Science and Technology "MISIS", 119049, Russia, Moscow, Leninskiy ave., 4

E-mail: chuvildeev@nifti.unn.ru

**Abstract**

The research aims to analyze the impact that severe plastic deformation arising during Rotary Swaging has on mechanical properties and corrosion resistance of a near-α-titanium alloy Ti-2.5Al-2.6Zr (Russian industrial name PT7M). The nature of corrosion decay in fine-grained alloys caused by hot salt corrosion is known to vary from pit corrosion to intercrystalline corrosion at the onset of recrystallization processes. Resistance to hot salt corrosion in a fine-grained titanium alloy Ti-2.5Al-2.6Zr is shown to depend on the structural-phase state of grain boundaries that varies during their migration as a result of "covering" corrosive doping elements (aluminum, zirconium) distributed in the crystal lattice of a titanium alloy.

---

(*) Corresponding author (nokhrin@nifti.unn.ru)

**Keywords:** Titanium alloys; fine-grained structure; Rotary Swaging; strength; hot salt corrosion; grain boundary; diffusion.

**Introduction**

Near-α-titanium alloys are extensively used today in nuclear engineering and nuclear power industry to produce heat exchange equipment for nuclear power plant (NPP) secondary systems [1, 2]. Such titanium alloys for NPPs shall comply with stringent requirements as to strength, ductility, and corrosion resistance at room and elevated operating temperatures [1-5]. To improve characteristics of titanium alloys, technologies based on optimization of their composition and different types of heat treatment and/or plastic deformation modes are now widely used [2-14].

Note that hot salt corrosion (HSC) is considered to be one of the most dangerous types of decay in titanium alloys [15-21]. It occurs at temperatures above 200-250°C when titanium interacts with halide salts (chlorides, bromides, iodides) and provided there is some water that can be bound into salt crystalline hydrates occluded in salt crystals or present in air. It shall be noted that while operating NPPs on ships or icebreakers, there might be cases when seawater or brine from a water desalination plant gets into a condensate-feeding system, which may lead to water-insoluble porous salt deposits being formed on the surface of heat-exchange tubes of steam generating units and to further concentration of corrosive components in those deposits: chlorides and bromides of alkali metals and alkaline earth metals. This gives rise to salt deposits formed in pores, which factor in HSC in heat-exchange tubing.

According to [19], resistance of near-α-titanium alloys to HSC is largely affected by the structural-phase state of grain boundaries, and formation of a fine-grained (FG) structure through severe plastic deformation (SPD) ensures a simultaneous increase in strength and corrosion resistance. The said approach involves Equal Channel Angular Pressing (ECAP) [22, 23] that guarantees fine fragmentation of the grain structure and diffusion redistribution of corrosive doping elements along grain boundaries of a titanium α-alloy Ti–(3.5-5)wt.%Al–(1.2-2.5)wt.%V (Russian industrial alloy PT3V).

Sticking to the above approach, this research aims to enhance strength and corrosion resistance of near-α-alloys Ti-(1.8-2.5)%Al-(2-3)%Zr (Russian industrial alloy PT7M) that are characterized by high heat resistance and are extensively used to manufacture modern NPPs.

**Materials and methods**

The target of research is a near-α industrial alloy PT7M with Ti-2.5wt.%Al-2.6wt.%Zr composition. The concentration of oxygen, nitrogen, hydrogen, and carbon in the alloy is 0.12 wt. %, 0.003 wt. %, 0.001 wt. % and 0.028 wt.%, respectively. The chemical composition of the alloy complies with Russian standard GOST 19807-91 requirements.

A fine-grained structure was formed with Rotary Swaging (RS) [24-26] using HF5-4-21 HMP machine (Germany). RS was performed at room temperature by means of gradual deformation of ⌀20 mm rod into ⌀16 mm rod (stage 1), ⌀12 mm rod (stage 2), ⌀10 mm rod (stage 3), ⌀8 mm rod (stage 4), and ⌀6 mm rod (stage 5). The deformation rate was 0.5-1 s$^{-1}$. The process of deformation involved high-speed action of hard-alloy bolts against the surface of a titanium rod with simultaneous axial rotation of the rod.

Structure studies were conducted using Leica IM DM metallographic microscope, Jeol JEM-2100 transmission electron microscope with JED-2300 energy dispersive X-ray analyzer and Jeol JSM-6490 scanning electron microscope with Oxford Instruments INCA 350 energy dispersive (EDS) microanalyzer. X-ray diffraction (XRD) analysis was performed with DRON-3 diffractometer (CuK$_α$ – radiation with wave length λ = 1.54178 Å, angular range 2Θ = 20-100°, sampling interval 0.02°, holding time at point 0.6 s). Diffraction patterns were analyzed using PhasanX 2.0 software (Powder Diffraction Phase Analysis, ver. 2.03). The scanning rate for overview XRD images was 2 °/min; for analyzing peak broadening the rate was 0.2 °/min. To offset the instrumentation interference $b_i$ from the installation into the broadening of XRD peaks, additional measurements of dependence of broadening on diffraction angle were taken on a reference specimen Silicon powder 99% 325 mesh. Intrinsic broadening of an XRD peak (hkl) was determined using the formula: $\beta_{hkl} = \sqrt{A_{hkl}^2 - b_{hkl}^2}$, where $A_{hkl}$ – experimentally obtained half-width of XRD peaks. The value of internal stresses $\sigma_{int}$ was calculated with Williamson-Hall method based on the inclination angle of $(\beta_{hkl})^2 \cdot \cos^2\Theta_{hkl}$ – $16\sin^2\Theta_{hkl}$ dependence [27], where $\Theta_{hkl}$ – angle corresponding to maximum intensity of an XRD peak indexed (hkl).

In order to test mechanical properties, Tinius Olsen H25K-S testing machine was used to perform tensile tests on cylindrical specimens with a working area of 3 mm in diameter. Microhardness was measured with HVS-1000 hardness tester under 200 g.

Autoclave HSC tests were carried out in a mixture of NaCl and KBr salts at the ratio of 300:1 at 250 ºC [19] with oxygen. Corrosion electrochemical studies were performed using R-8 potentiostat-galvanostat (Russia) at room temperature in an aqueous solution of 0.2%HF+10%HNO$_3$. Based on the analysis of Tafel portions of potentiodynamic dependences 'potential $E$ – current density $i$' in semilogarithmic coordinates lg($i$) - $E$, corrosion current density $i_{cor}$ and corrosion potential $E_{cor}$ were identified. The surface of specimens prior to corrosion tests was mechanically polished to reach the level of roughness of 3-5 μm. Ahead of electrochemical tests, the surface of the specimen was covered with corrosion-resistant coating except for an area of 5×10 mm$^2$ (corrosive) located either in the center (variant 1) or in the subsurface (variant 2) of the cross-section of a specimen. Prior to electrochemical tests, the specimen was kept in an electrochemical cell in an aqueous solution of 0.2%HF+10%HNO$_3$ until a stationary potential value was reached, after which $E_{cor}$ ($i_{cor}$) dependence was surveyed at the scanning rate of 0.5 mV/s.

**Results and discussion**

A near-α-alloy Ti-2.5Al-2.6Zr in the initial state has an inhomogeneous coarse-grained plate-like and needle-like structure with β-phase particles stitched along the boundaries of titanium α'-phase (see Fig.1a, b) typical of α- and near-α titanium alloys [1-4, 28]. Plates of titanium α'-phase in the initial state are ~5-10 μm deep. The conclusion that particles in Fig. 1b are β-phase has been made based on EBSD analysis since the particles have a BCC [1] and on literature values [29, 30].

After RS, a fine-grained subgrain structure with an average fragment size of ~0.2-0.5 μm is formed in the alloy structure (see Fig.2a-2d). EDS results show that the boundaries of fine-grained alloy fragments are zirconium-rich – local concentration of Zr atoms in the crystal lattice is ~1.6-2.0 wt.%, while the amount of zirconium along fragment boundaries approximates 3.1-3.6 wt.% (Fig. 3). An increase in the amount of zirconium in fragment boundaries triggers competitive segregation – aluminum concentration in fragment boundaries is ~1.4-1.8 wt.%, which is somewhat less than the amount of aluminum in the grain volume (~2.1-2.3 wt.%). The results of electron microscopic studies prove that the parameters of the crystal structure of particles along grain boundaries of PT7M alloy are close to β-phase parameters.

Fig. 4 shows XRD findings for Ti-2.5Al-2.6Zr near-α-alloy specimens in various structural states. A coarse-grained alloy has all the major peaks corresponding to α(α')-phase[2] (see Fig. 4a). XRD analysis reveals no β-phase particles in the parent coarse-grained alloy, which obviously relates to their small volume fraction.

XRD analysis proves that RS causes broadening of α-Ti peaks and their slight shift towards the area of small reflection angles – diffraction angle $2\Theta_{max}$ for α-Ti XRD maximum (100) in Ti-2.5Al-2.6Zr alloy after RS increases from 35.11º to 35.05-35.07º, while its half-width at half-height ($\beta_{0.5}$) goes up from 0.308º to 0.554º. A shift of XRD peaks towards the area of small diffraction angles in line with Wulff–Bragg's condition ($2d_{hkl} \cdot \sin\Theta_{max} = n\lambda$, where $d_{hkl}$ – interplanar spacing, $\lambda = 1.54178$ Å – X-ray wavelength) indicates a decrease in interplanar spacing $d_{hkl}$ and formation of compressive internal stresses ($\sigma_{int}$) after RS. The stress value $\sigma_{int}$ determined with Williamson-Hall method by the inclination angle of $(\beta_{0.5})^2 \cdot \cos^2\Theta_{max} - 16\sin^2\Theta_{max}$ dependence [27] (see Fig. 5) amounts to 160 MPa for a coarse-grained alloy and 440-520 MPa for a near-α-alloy Ti-2.5Al-2.6Zr after RS.

---

[1] Note that since it is highly complicated to identify the structure of submicron particles of these phases (including with electron microscopy), there is no consensus among researchers about the nature of these particles. In particular, PT7M alloy developers believe that these particles are β-phase [5, 29, 30]. At the same time, references [14,31] focused on studying the structure of near-alloy PT3V (Ti-4Al-2V) with TEM prove that some of these particles can be α″-phase or $α_2$-phase particles.

[2] Due to identical crystallographic structure, XRD peaks of α- and α'-phase in titanium XRD patterns are indistinguishable.

Hardness $H_v$ of a fine-grained alloy in the cross-section goes slightly down from 2.9-3.0 GPa to 2.5-2.6 GPa when moving from the edge to the center of the cross-section (hardness of an alloy in the original coarse-grained state approximates ~1.9-2.0 GPa). Tensile tests show that formation of a fine-grained structure leads to an increase in yield strength $\sigma_{0.2}$ and ultimate strength $\sigma_b$ from 280 MPa to 1050-1070 MPa and from 590 MPa to 1080 MPa, respectively. Meanwhile tensile elongation ($\delta_5$) drops from 40% to 6-8%. Deformation curves $\sigma(\varepsilon)$ for coarse-grained and fine-grained alloy specimens are provided in Fig. 6.

Analysis of data obtained from fractographic studies of fractures after tensile tests of Ti-2.5Al-2.6Zr alloy specimens shows that the fracture in a coarse-grained alloy specimen (Fig.7a) is viscous with typical zones of slow and fast crack propagation, as well as the break zone that are considered standard for a titanium α-alloy. According to the classification provided in [29], the fracture in the zone of slow (initial) crack propagation has the form of equiaxial pits (Fig.7b).

The surface of a fracture in a fine-grained specimen after RS consists of three zones: fibrous zone (zone 1 in Fig.7c) corresponding to the area of slow crack propagation, radial zone (zone 2 in Fig.7c) corresponding to the area of fast unstable crack propagation, and break zone (zone 3 in Fig.7c). It is noteworthy that the radial zone (zone 2) is unusually spiral, which is obviously due to deformation processing during RS – dynamic deformation of the surface with hard-alloy bolts with simultaneous axial rotation of a specimen. According to the classification provided in [29], the fracture in the area of slow crack propagation has the form of equiaxial pits that characterize viscous intracrystalline decay (Fig.7d), while the fracture in the radial area is characterized by a number of pits set circumferentially (Fig.7e), which is caused by the specific nature of plastic deformation during RS. The break zone in a fine-grained Ti-2.5Al-2.6Zr alloy specimen in line with the classification provided in [29] can be characterized as a set of shear pits (Fig.7f).

Note that in general the fracture of a fine-grained specimen is typically viscous. Thus, a significant decrease in the plasticity of Ti-2.5Al-2.6Zr alloy after RS can be primarily attributed to the formation of tensile stresses (see internal stress calculation based on XRD data).

HSC tests reveal two types of corrosion defects observed in a coarse-grained near-α-alloy Ti-2.5Al-2.6Zr. Corrosion defects of the first type are typically stitched (Fig. 8a), which is associated with a stitched arrangement of β-phase particles in the structure of the parent alloy (Fig. 1a)[3]. Corrosion defects of the second type are typical of intercrystalline corrosion (ICC) (Fig. 8b). The corroded layer in a coarse-grained alloy is 200-300 μm deep.

A fine-grained Ti-2.5Al-2.6Zr alloy has a different corrosion decay pattern – after 270 h of testing conducted at 250 °C, pit corrosion is observed in a fine-grained near-α-alloy Ti-2.5Al-2.6Zr (Fig. 9a). Maximum depth of corrosion defects is under 70 μm. When testing time reaches 320 h, apart from pit corrosion the surface of a fine-grained alloy shows traces of ICC (Fig. 9b). When testing time is increased to 520 h, ICC defects are several times deeper than pit corrosion defects (Fig. 9c).

---

[3] The investigated alloy Ti-2.5Al-2.6V belongs to the so-called near-α titanium alloys where a volume fraction of β-phase particles is small (under 5 vol.%). This volume fraction is insufficient to form a continuous layer of β-phase particles around α-phase boundaries, therefore β-phase particles in near-α titanium alloys are arranged as separate micron and submicron particles along α- or α′-phase boundaries (see Fig.1b). This essentially distinguishes near-α titanium alloys from a two-phase (α+β)-alloy Ti6Al4V that has a volume fraction of β-phase particles totaling 15-20%. Thus, an almost completely closed system of interphase boundaries is formed. The impact of single β-phase particles on the susceptibility of a near-α titanium alloy to ICC is rather unexpected. We reckon that it might relate to the effect known as complete and/or incomplete wetting of grain boundaries with the second solid phase. It was frequently observed in the last years and can influence the studied processes and can influence the studied processes and the susceptibility of near-α titanium alloys to ICC.

The surface of specimens of coarse-grained and fine-grained alloys Ti-2.5Al-2.6Zr after HSC tests shows porous salt deposits of various composition varying in color from dark grey to ochroleucous. The formation of pores in salt deposits observed on the surface of specimens indicates gaseous $TiCl_4$ formed during HSC [16]. Based on XRD findings, corrosion on the surface of specimens is a mixture of NaCl (PhasanX 2.0 Card No.01-0993) and $TiO_2$ in two polymorphic modifications (rutile – Card No.73-1764, anatase – Card No.73-1765) (Fig. 10a).

XRD patterns of near-α-alloy Ti-2.5Al-2.6Zr after HSC show no peaks corresponding to β-phases (Fig. 10b). XRD analysis proves that after HSC testing at 250 ºC for 520 h, $\sigma_{int}$ calculated with Williamson-Hall method goes down from 440-520 MPa to 110-150 MPa and appears to be close to the level of internal stresses in near-α-alloy Ti-2.5Al-2.6Zr in the initial state (Fig. 4).

We reckon that any changes in the nature of corrosion decay in a fine-grained alloy triggered by longer testing times relate to the onset of recrystallization leading to grain growth. According to [19], intensively migrating grain boundaries in fine-grained titanium α-alloys bring forth atoms of corrosive doping elements from the crystal lattice. An increase in their local concentration along grain boundaries leads to an increase in electrochemical potential difference 'crystal lattice – grain boundary', and therefore to an increase in ICC rate for fine-grained titanium α-alloys[4].

---

[4] Note that one of the reasons for changes in the corrosion mechanisms might be the above effect known as incomplete wetting of grain boundaries. Formation of a UFG structure in a near-α titanium alloy is expected to significantly increase the area of grain boundaries and consequently eliminate the wetting of grain boundaries. Eventually, the susceptibility of a titanium alloy to ICC shall decrease while its corrosion resistance shall grow. Grain growth during HSC tests accompanied by a decrease in the area of grain boundaries may again contribute to complete or incomplete wetting of grain boundaries, formation of a partially closed system of interphase boundaries, and therefore to the above effect known as the corrosion decay mechanism shift from pit corrosion to ICC.

This assumption is in good agreement with the results of the research into the structure and properties of a fine-grained alloy Ti-2.5Al-2.6Zr after HSC tests.

The structure of a fine-grained alloy specimen after testing for 270 h looks typical for primary recrystallization – coarse recrystallized grains amid a highly deformed matrix (Fig.11a, b). The average size of coarse grains in a partially recrystallized structure of a specimen is ~ 8.2 µm for surface layers and ~9.6 µm for the central part of the cross-section. Smaller grain size in the surface of a rod apparently relates to its bigger deformation, which consequently leads to an increase in the number of recrystallization nuclei, smaller distance between them, and consequently, to a decrease in the average size of recrystallized grains. Microhardness of the surface and the central part of the cross-section in Ti-2.5Al-2.6Zr alloy specimens after 270 h of HSC testing at 250 ºC is 2.2 GPa and 2.0-2.1 GPa, respectively. This is notably lower than microhardness of Ti-2.5Al-2.6Zr alloy after RS ($H_v$ = 2.9-3.0 GPa – for the surface; $H_v$ = 2.5-2.6 GPa – for the central part of a rod).

Structure studies show that after testing for 320 h and 520 h, the average grain size in a fine-grained alloy is ~3.5 µm and 4.2 µm, respectively. The alloy structure after holding for 520 h is fully recrystallized (Fig. 11c, d). The average microhardness of specimens is 1.8 GPa and 1.7 GPa, respectively. Note that microhardness in the central part of the cross-section is 100-150 MPa smaller than the microhardness of the surface layer, while the average grain size of the surface layer appears to be 0.5-1 µm smaller than the average grain size in the central part of the cross-section. (Fig. 11c, d). Inhomogeneity of the grain structure in a recrystallized specimen is clearly visible in the macro section[5] showing the area of particular etch ability of a fine-grained structure in the central part of a rod (Fig. 12).

An indirect proof of growing concentration of doping elements along migrating grain boundaries might be an increase in the Hall-Petch coefficient in recrystallized alloys – Fig.13 shows that the inclination angle of the dependence of microhardness on the average grain size in $H_v - d^{-1/2}$ coordinates amid coarse grains is noticeably larger than amid small grains. As shown in [30-32], grain boundary hardening coefficient K in the Hall-Petch relation ($H_v = H_{v0} + Kd^{-1/2}$) that characterizes the level of grain boundaries resistance to lattice dislocations passing through them heavily depends on the structural state of grain boundaries and may grow substantially in case of grain boundary segregations: in pure iron K = 0.5 MPam$^{1/2}$, in iron alloyed with carbon K = 1.6 MPam$^{1/2}$ [30]. A similar effect for ultrafine-grained α-titanium alloys obtained by Equal Channel Angular Pressing was described in [19]. Based on these assumptions, high value of K in a recrystallized alloy Ti-2.5Al-2.6Zr after prolonged HSC testing might result from high concentration of atoms of doping elements along grain boundaries. This may lead to grain boundary hardening, but at the same time, segregations along grain boundaries dramatically reduce corrosion resistance in titanium alloys [19, 33].

---

[5] Etching in an aqueous solution of 15%HF+10%HNO$_3$ + 35% glycerine

Dependences "potential $E$ – holding time in electrolyte $t$" and "potential $E$ – current density $i$" for coarse-grained and fine-grained alloy specimens are shown in Fig. 14. The results of electrochemical studies show that while forming a fine-grained structure, the corrosion current density ($i_{cor}$) for the metal in the central part of the cross-section is 1.14-1.16 mA/cm$^2$. At the same time, corrosion potential $E_{cor}$ for the material in the central part of the specimen amounts to –(472-485) mV. Corrosion current density on the surface of a fine-grained alloy specimen after RS is $i_{cor}$ = 0.80-0.94 mA/cm$^2$ ($E_{cor}$ = –(483÷492) mV). Thus, a severely deformed surface layer of a Ti-2.5Al-2.6Zr alloy specimen is characterized by higher corrosion resistance (lower corrosion current density $i_{cor}$ and lower corrosion potential $E_{cor}$) as compared to the central part of a specimen characterized by lower hardness. The average corrosion current density for a coarse-grained Ti-2.5Al-2.6Zr alloy is $i_{cor}$ = 1.07 mA/cm$^2$ ($E_{cor}$ = –465 mV).

Note that higher corrosion resistance in the surface layer of a rod is rather unexpected since it is traditionally assumed that stronger deformation leads to faster corrosion decay [34-37]. This is indirectly proven by the studies into the microstructure of titanium alloy rods – Fig.12 shows that the central part of a rod after RS is subjected to more intensive etching (corrosion) as compared to the surface of a processed material.

Metallographic studies of the surface of specimens after electrochemical investigation show that corrosion while testing a coarse-grained alloy takes place predominantly due to etching of interphase (α-β)-boundaries (Fig. 15a, b). The nature of corrosion damage to the surface of a fine-grained specimen is different and is characterized by etching localized deformation bands (Fig. 15c, d).

Based on the data presented in Fig. 15, we reckon that the observed effect[6] can be explained by the fact that localized deformation bands in the surface layer of a titanium rod are located much closer to each other almost overlapping at the surface. As a result, the metal structure may have no alternating severely and lightly deformed structures with different electrochemical properties that are somewhat analogous to micro-galvanic couples triggering faster metal dissolution.

The second factor that may contribute to better corrosion resistance of deformed areas in a titanium alloy is a change in electrochemical conditions of corrosion at the interface between α- and β-phases. β-phase of titanium is known to have a more negative electrode potential ($E_β$) than α-phase ($E_α$), which among other things leads to its intense etching during corrosion tests (Fig.1a, 15a, b) and a shift of corrosion potential $E_{cor}$ in two-phase α+β-alloys (Ti6Al4V, Russian industrial alloy VT6, etc.) towards more negative corrosion potentials as compared to pure titanium [38].

Plastic deformation of grains leading to increased density of lattice dislocations according to [38, 39] is expected to reduce electrode potential of α-phase (shift potential $E_α$ towards more negative values). This means that the potential difference between α- and β-phase ($ΔE = E_α - E_β$) will decrease and therefore reduce the contribution of interphase (α-β)-boundaries into the intensity of corrosion failure in a fine-grained titanium alloy. An additional factor that reduces the contribution of interphase (α-β)-boundaries is refining β-phase particles during plastic deformation.

---

[6] High corrosion resistance of surface layers in titanium rods with bigger prior deformation (as compared to the metal in the center of the cross-section).

The obtained result proves better corrosion resistance of a fine-grained Ti-2.5Al-2.6Zr alloy as compared to the original state, as well as changes in the mechanism of corrosion decay of alloy while forming a fine-grained grain-subgrain structure in it.

**Conclusions**

1. Rotary Swaging helps to form a fine-grained structure with enhanced mechanical properties in a titanium α-alloy Ti-2.5Al-2.6Zr (Russian industrial alloy PT7M) – yield strength (1050-1070 MPa), ultimate strength (1080 MPa), and hardness (2.9-3.0 GPa). After RS, tensile fields of internal stresses are formed in the alloy that reduce the alloy plasticity – tensile elongation to failure is 6-8%.

2. It is shown that there are two types of corrosion defects observed on the surface of a coarse-grained near-α-titanium alloy Ti-2.5Al-2.6Zr after HSC tests: defects related to stitched β-phase particles (defects of type I) and ICC defects (defects of type II) related to higher concentration of corrosive doping elements (zirconium, aluminum) along grain boundaries. The nature of HSC-driven decay that occurs in UFG alloys depends on testing time and changes from pit corrosion to ICC when testing time increases from 270 h to 500 h. Changes in the nature of corrosion decay affecting a UFG alloy Ti-2.5Al-2.6Zr occur when recrystallization starts, during which intensively migrating grain boundaries bring forth corrosive doping elements distributed in the crystal lattice of a titanium alloy Ti-2.5Al-2.6Zr. The possible reason for a shift in the corrosion decay mechanisms during HSC tests might be the effect of complete or incomplete wetting of grain boundaries with β-phase.

3. Formation of a fine-grained structure is found to increase the resistance of a titanium alloy to electrochemical corrosion, meanwhile differences in corrosion resistance between surface and central areas of the cross-section in Ti-2.5Al-2.6Zr alloy specimens are obviously caused by differences in the extent of cumulative deformation in the central and surface layers of a rod during RS. During electrochemical testing, corrosion damage to the surface of a coarse-grained specimen occurs through etching interphase (α-β)-boundaries; active corrosion etching of localized deformation bands is observed in a fine-grained alloy.

**Data Availability**

The raw/processed data required to reproduce these findings cannot be shared at this time as the data also forms part of an ongoing study.

**List of figures**

**Figure 1**. Structure of coarse-grained near-α-alloy Ti-2.5Al-2.6Zr (optical microscopy). β-phase particles marked with arrows (see fig. 1b).

**Figure 2**. Structure of fine-grained near-α-alloy Ti-2.5Al-2.6Zr after RS (TEM)

**Figure 3**. EDS analysis of the grain boundaries composition in fine-grained alloy Ti-2.5Al-2.6Zr after RS

**Figure 4**. XRD analysis of Ti-2.5Al-2.6Zr alloy in the original state (a) and in a fine-grained state after RS (b)

**Figure 5**. Williamson-Hall curves plotted after analyzing XRD patterns of Ti-2.5Al-2.6Zr alloy specimens in the original state (1), in a fine-grained state after RS (2) and in a fine-grained state after HSC testing for 520 h

**Figure 6**. Strain hardening curves plotted during tensile tests of coarse-grained (1) and fine-grained (2) Ti-2.5Al-2.6Zr alloy specimens

**Figure 7**. Fractographic analysis of factures in coarse-grained (a, b) and fine-grained (c, d, e, f) Ti-2.5Al-2.6Zr alloy specimens after tensile tests. Fig.7c: (1) – fibrous zone of slow crack propagation (Fig.7d); (2) – radial zone of fast unstable crack propagation (Fig.7e); (3) – break zone (Fig.7f). Scanning electron microscopy

**Figure 8**. Standard view of corrosion defects on the surface of a coarse-grained near-α-alloy Ti-2.5Al-2.6Zr after HSC tests: a) type I – 'stitched' corrosion defects; b) type II - ICC

**Figure 9**. Standard view of corrosion defects on the surface of a fine-grained near-α-alloy Ti-2.5Al-2.6Zr after HSC tests conducted for 270 h (a), 300 h (b) and 520 h (c)

**Figure 10**. XRD analysis results after HSC testing for 520 h: a) analysis of the phase composition of corrosion products on the specimen surface; b) analysis of the phase composition of the surface in a fine-grained alloy after HSC testing and mechanical corrosion removal

**Figure 11**. Microstructure of an edge (a, c) and central part (b, d) of the cross-section in fine-grained Ti-2.5Al-2.6Zr alloy specimens after HSC tests for 270 h (a, b) and 520 h (c, d). Interference contrast

**Figure 12**. Microstructure of the cross-section in ⌀6 mm Ti-2.5Al-2.6Zr alloy specimen after rotary swaging (a) and after HSC testing for 520 h (b)

**Figure 13**. Dependence of microhardness on grain size in $H_v - d^{-1/2}$ coordinates for fine-grained Ti-2.5Al-2.6Zr alloy specimens after HSC testing

**Figure 14**. Results of electrochemical studies of coarse-grained (1) and fine-grained (2) Ti-2.5Al-2.6Zr alloy specimens: a) $E(t)$ dependence; b) $E(i)$ dependence. Curve (2-1) – cross-section center in a fine-grained specimen, (2-2) – surface layer in a fine-grained specimen.

**Figure 15**. Structure of the surface in coarse-grained (a, b) and fine-grained (c, d) Ti-2.5Al-2.6Zr alloy specimens after electrochemical testing: c – surface layer in a fine-grained alloy; d – central layer in a fine-grained alloy

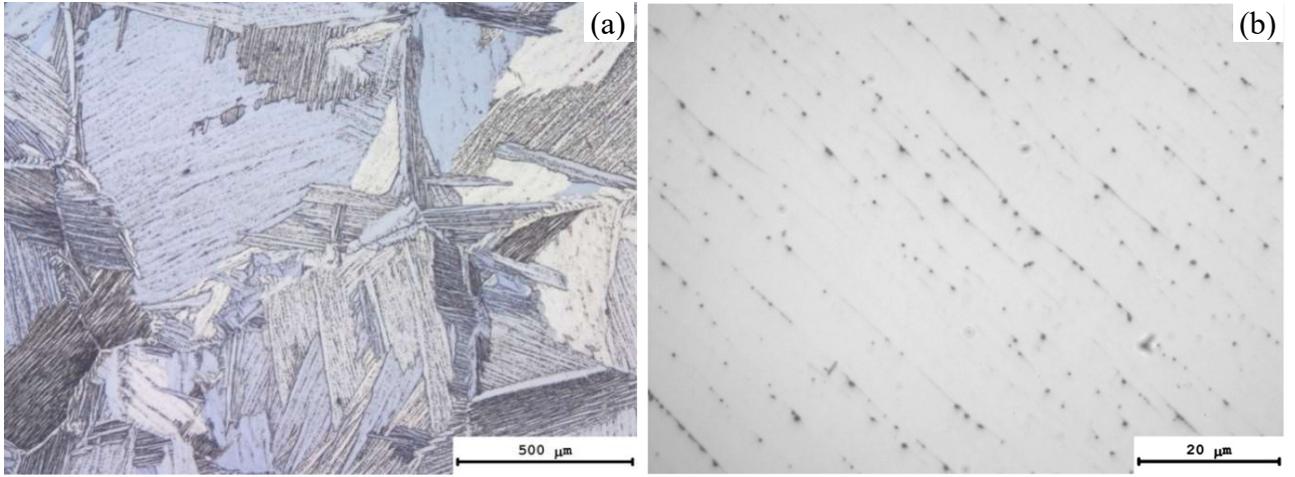

Figure 1

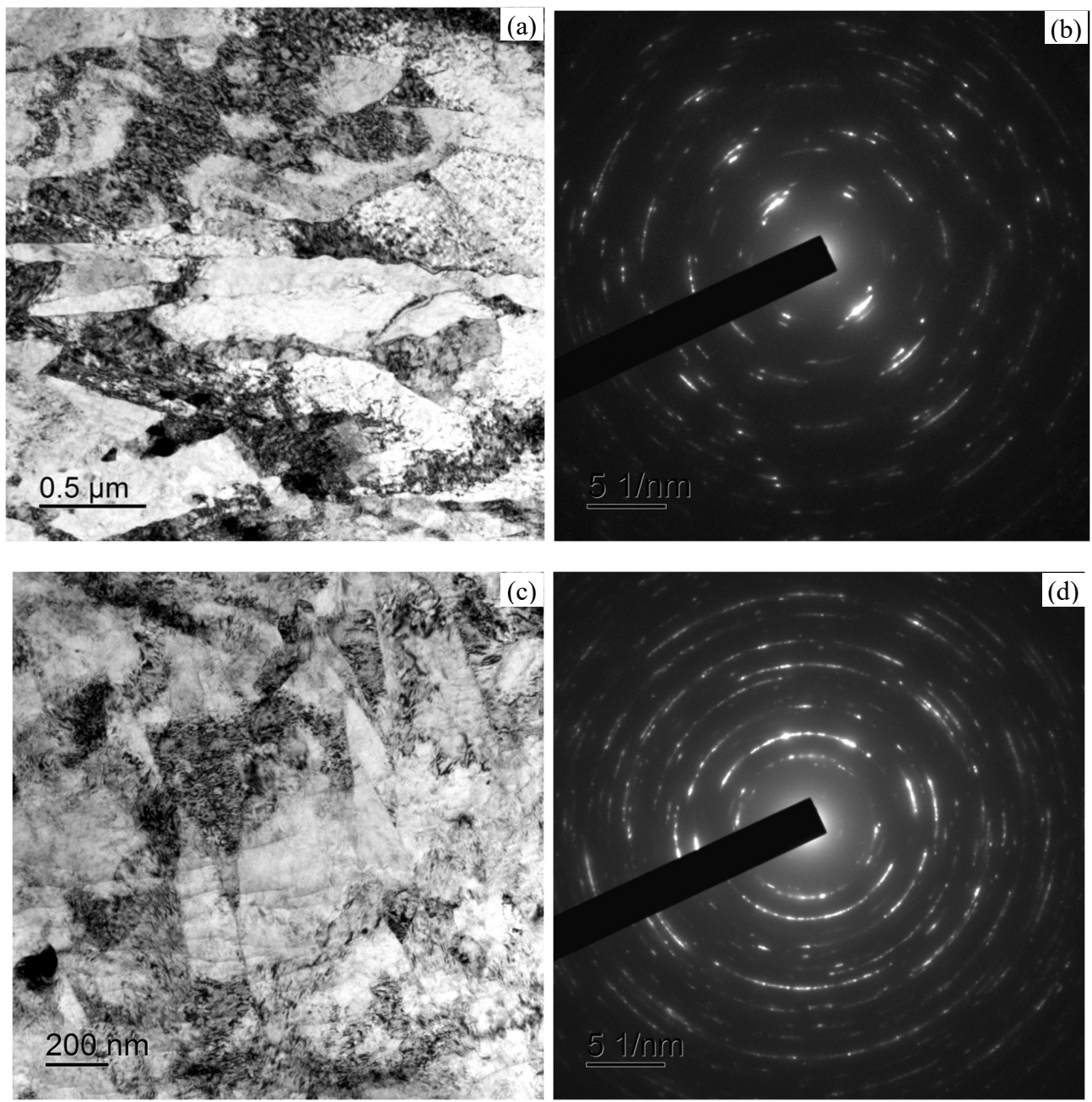

Figure 2

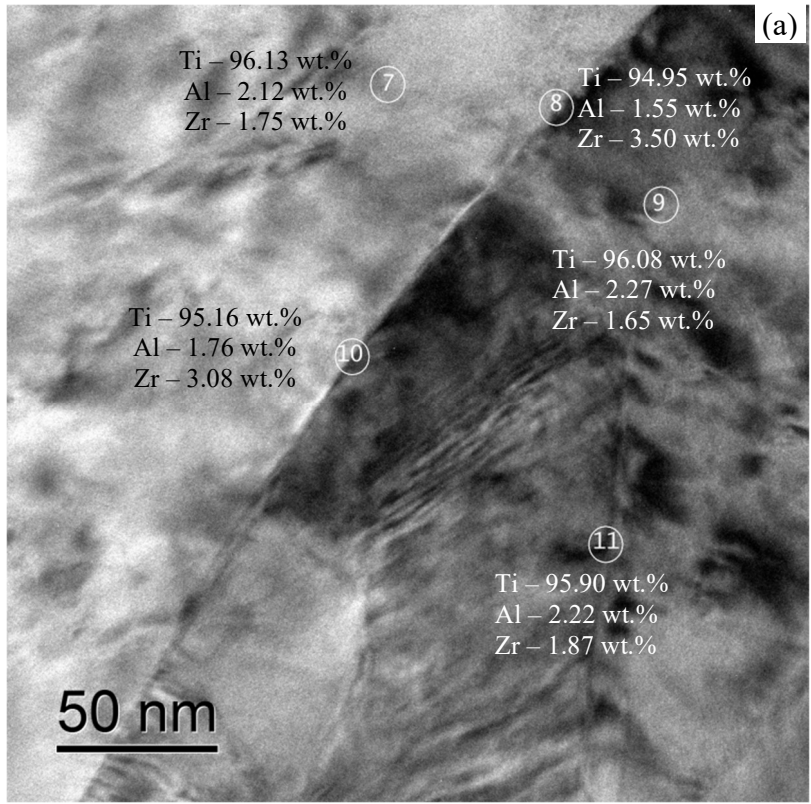

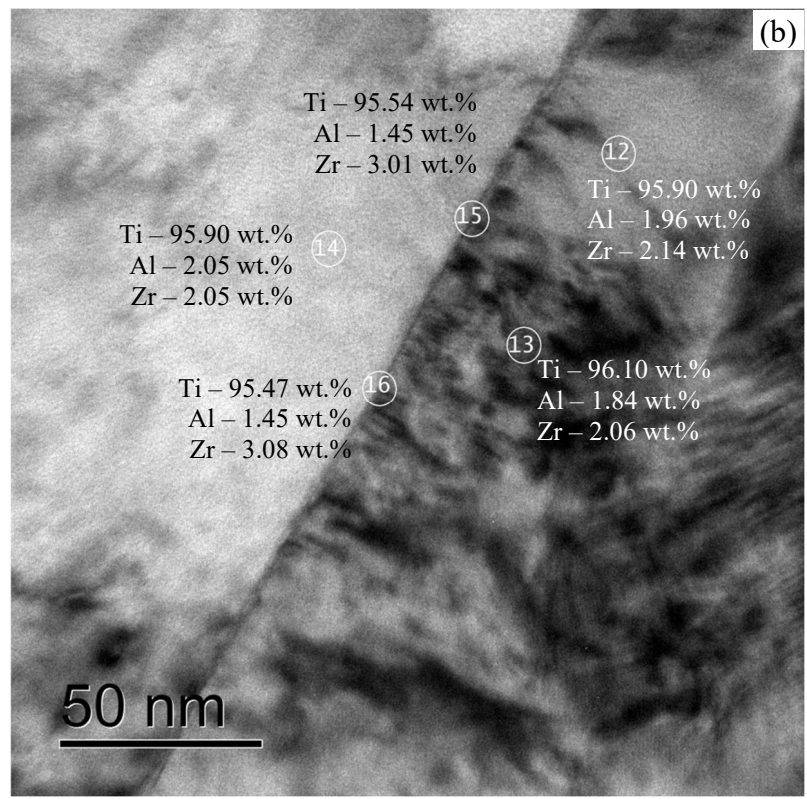

Figure 3

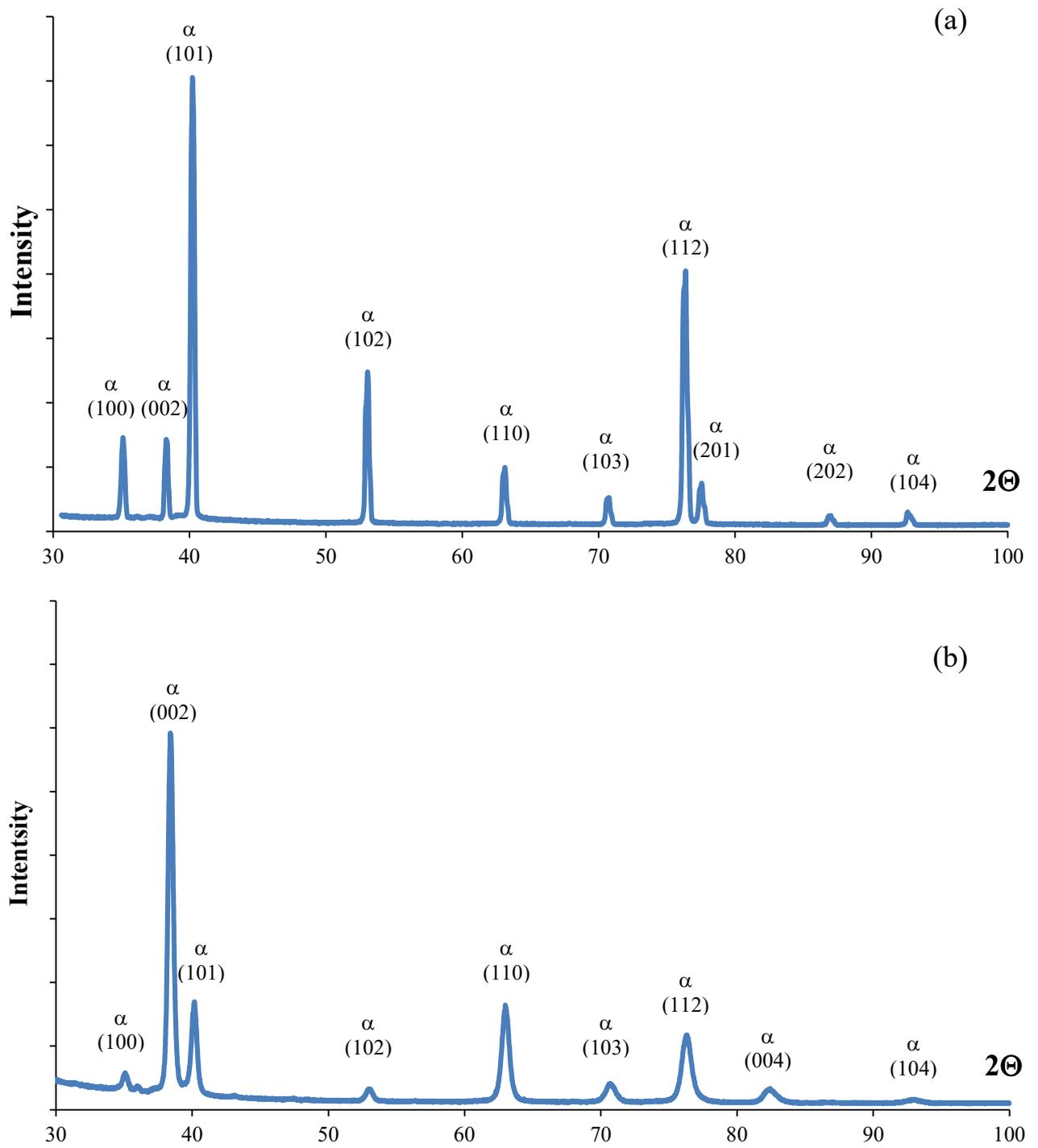

Figure 4

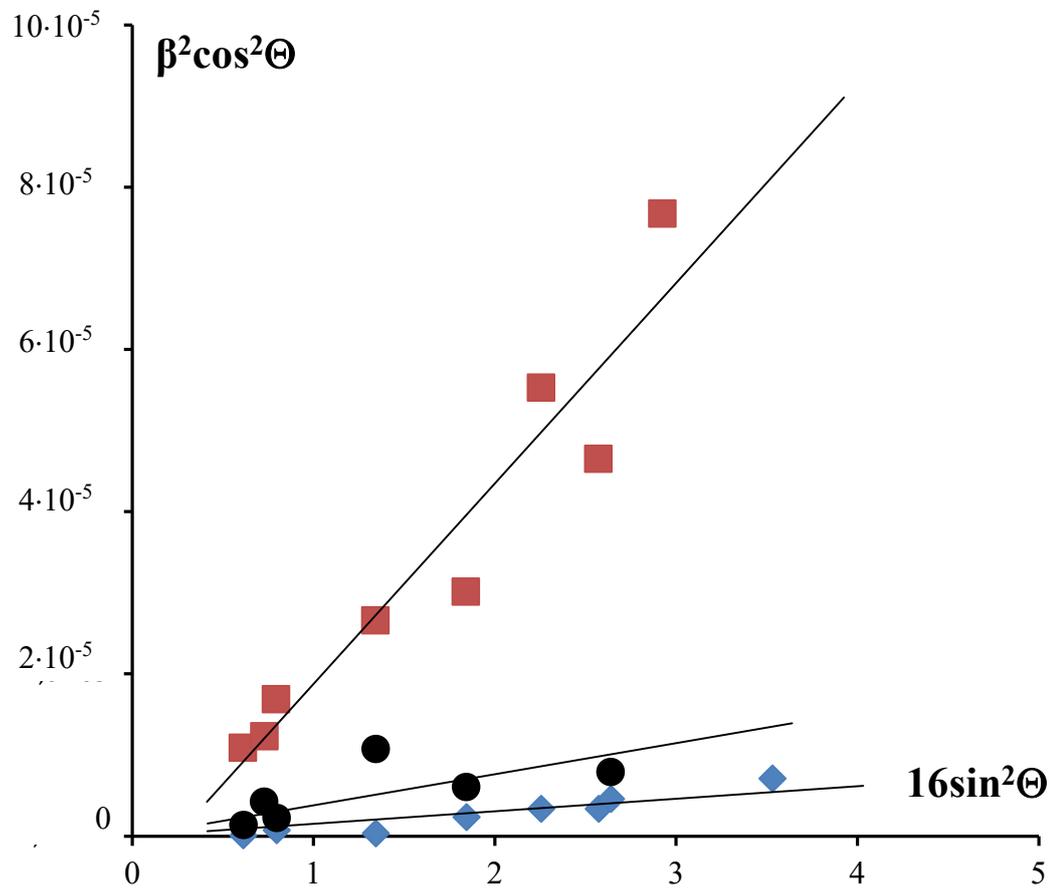

Figure 5

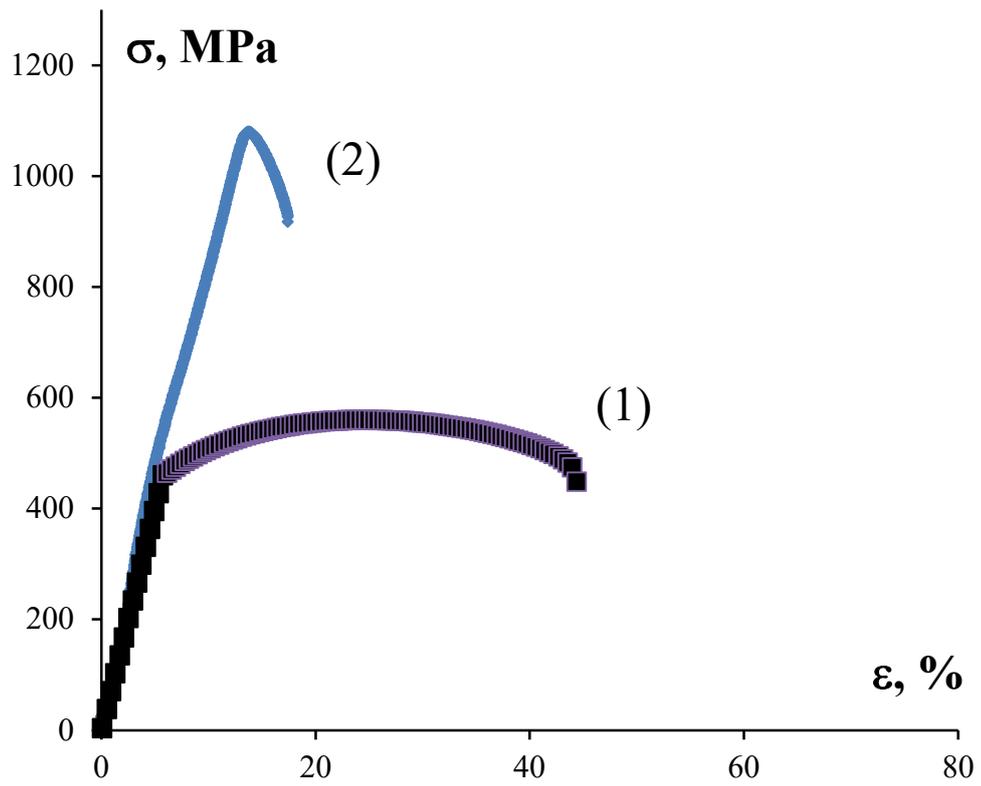

Figure 6

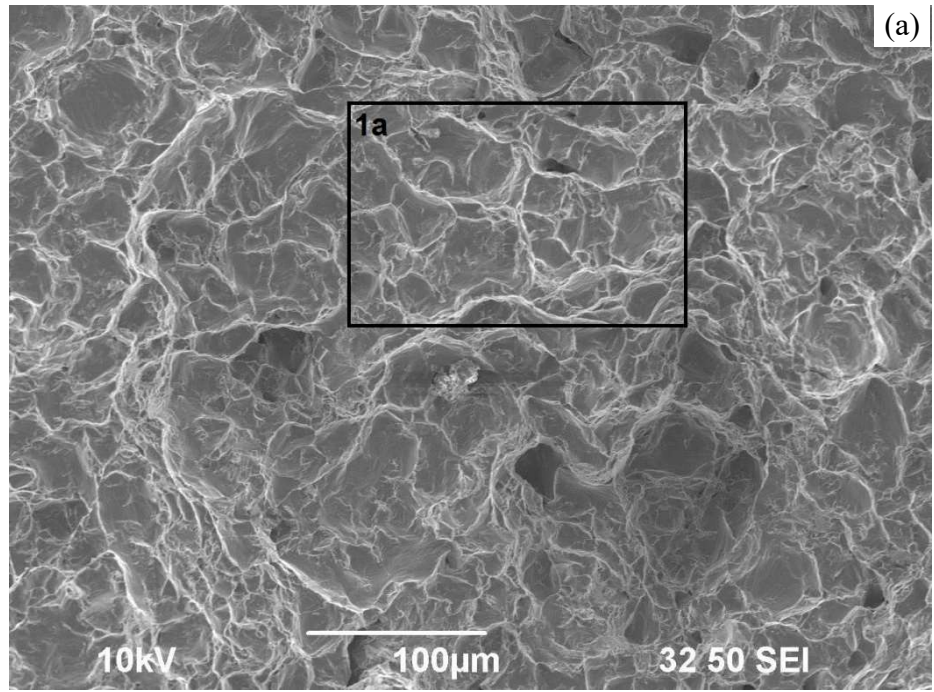

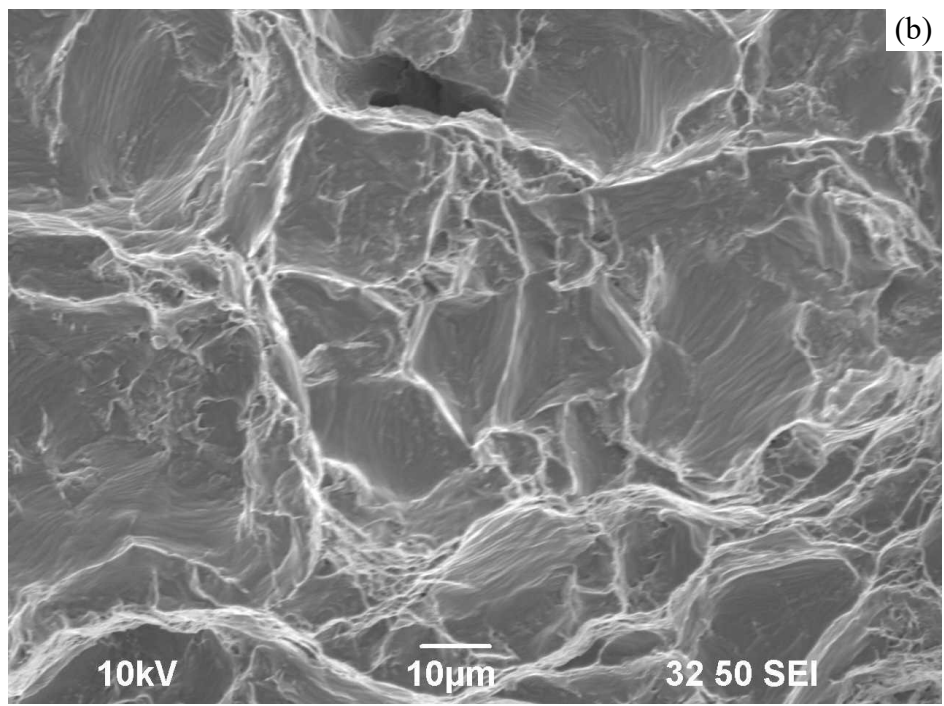

Figure 7 (starting)

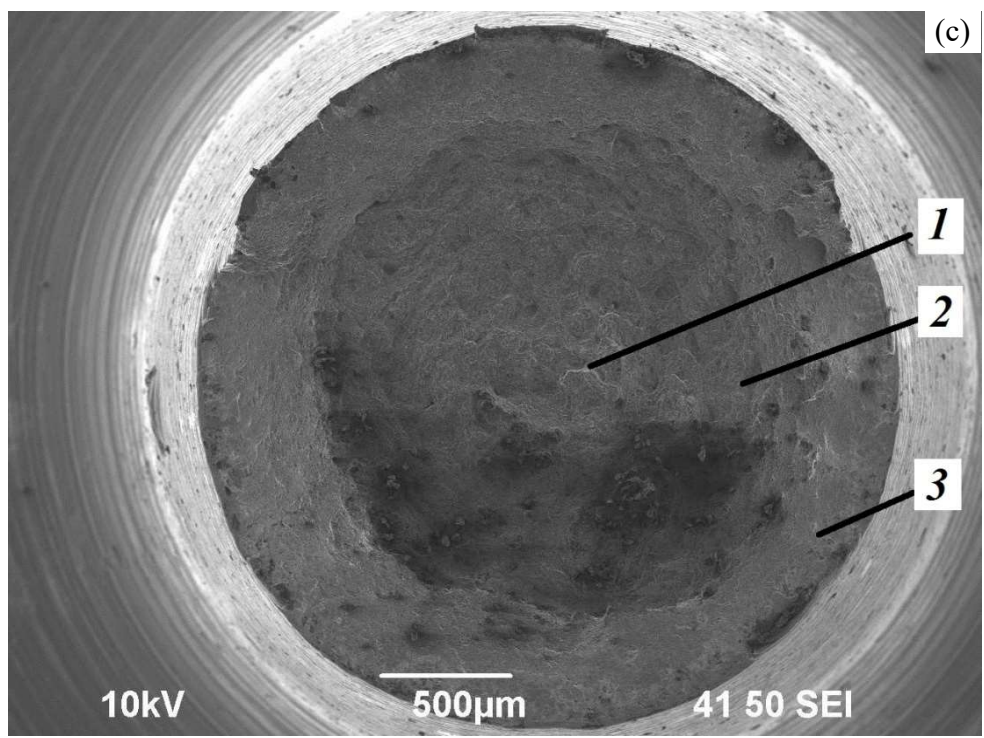

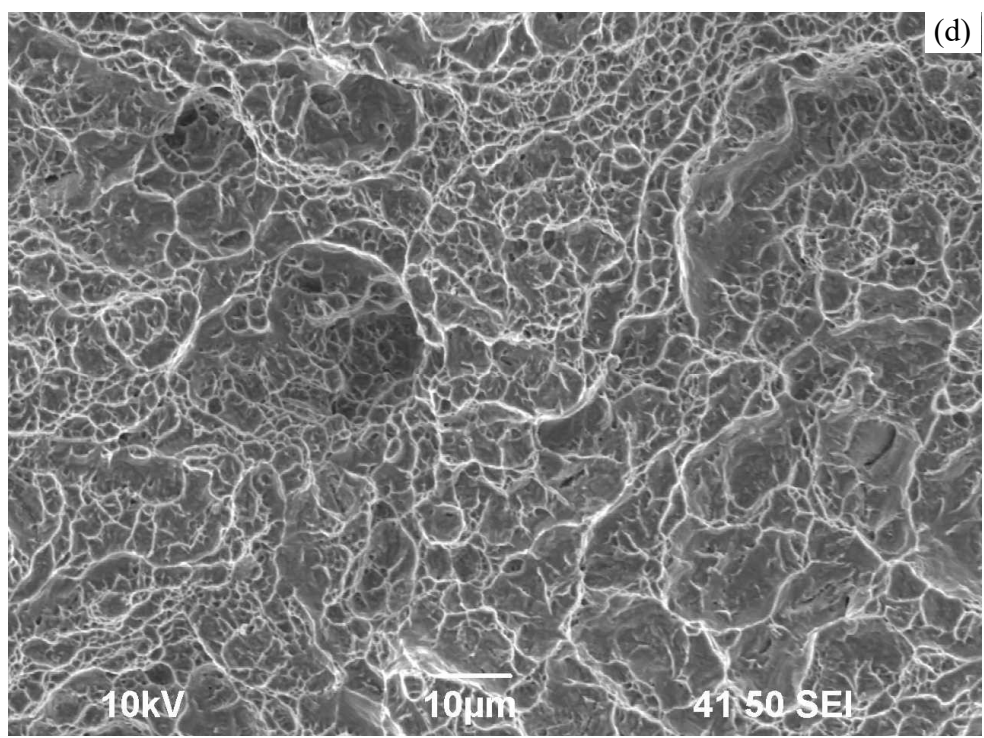

Figure 7 (continuation)

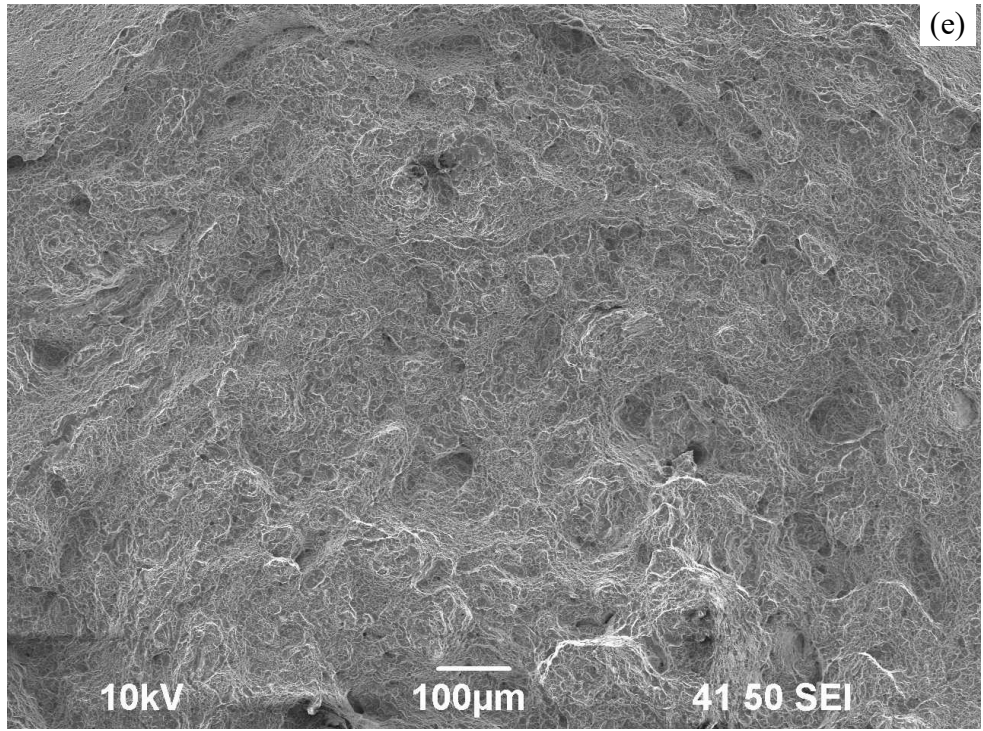
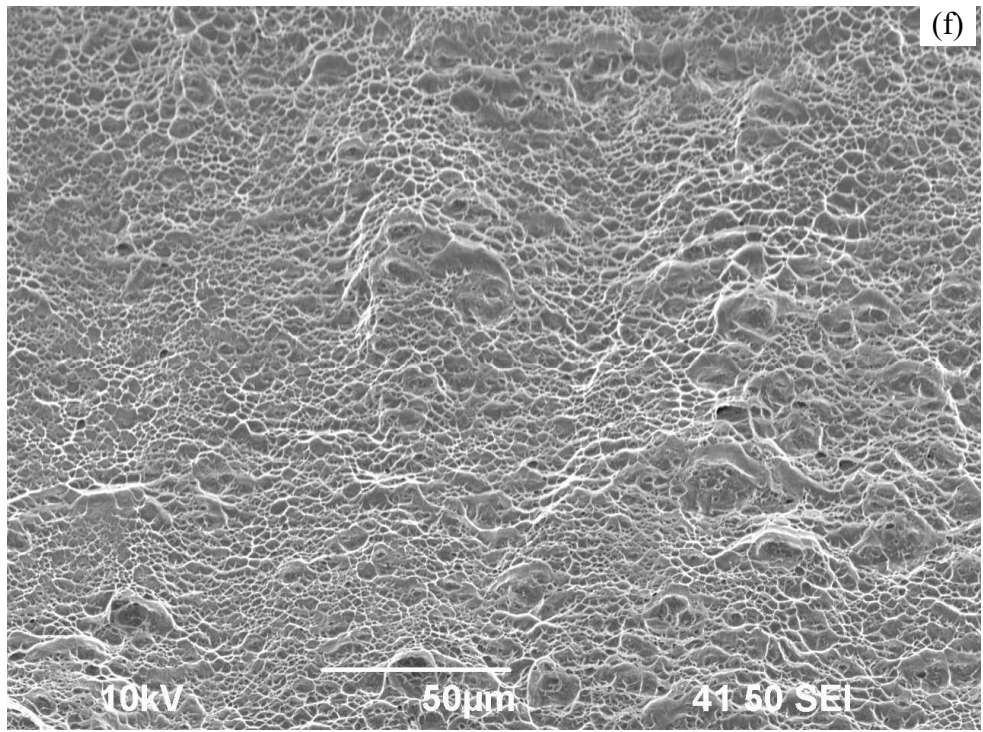

Figure 7 (end)

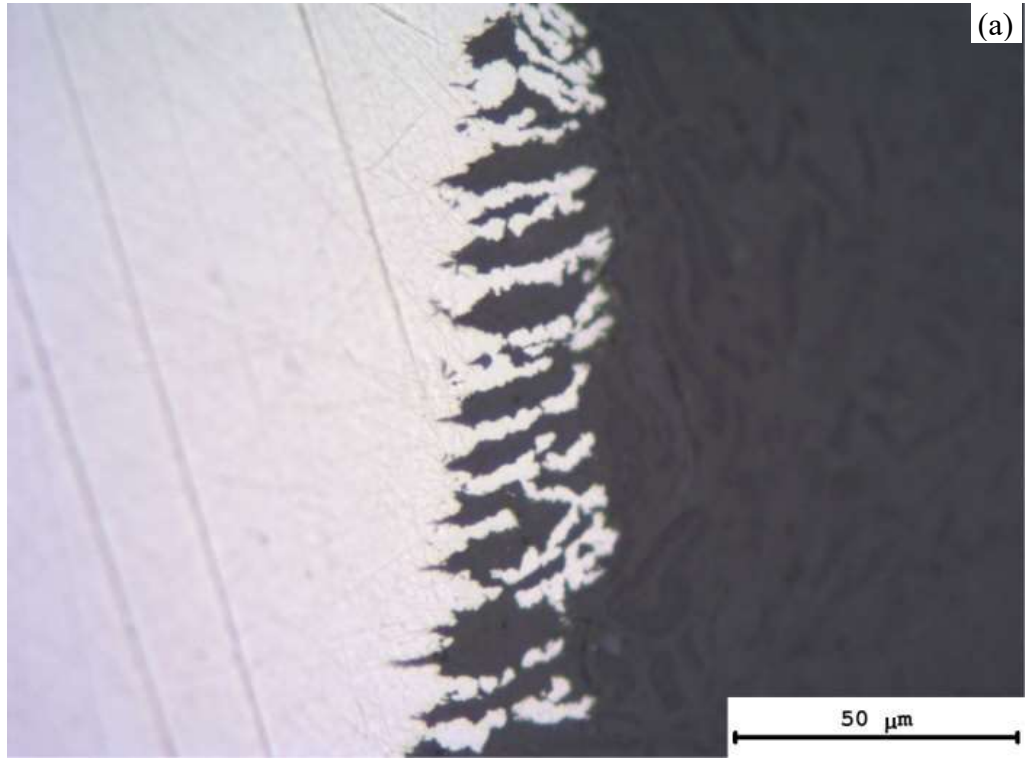
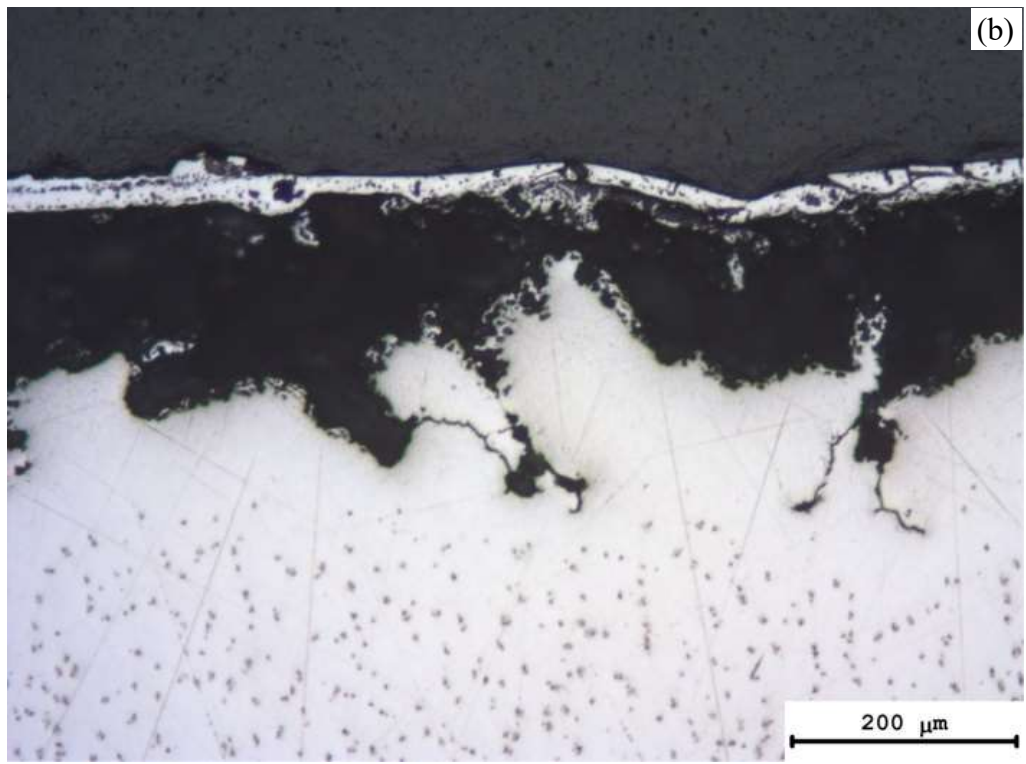

Figure 8

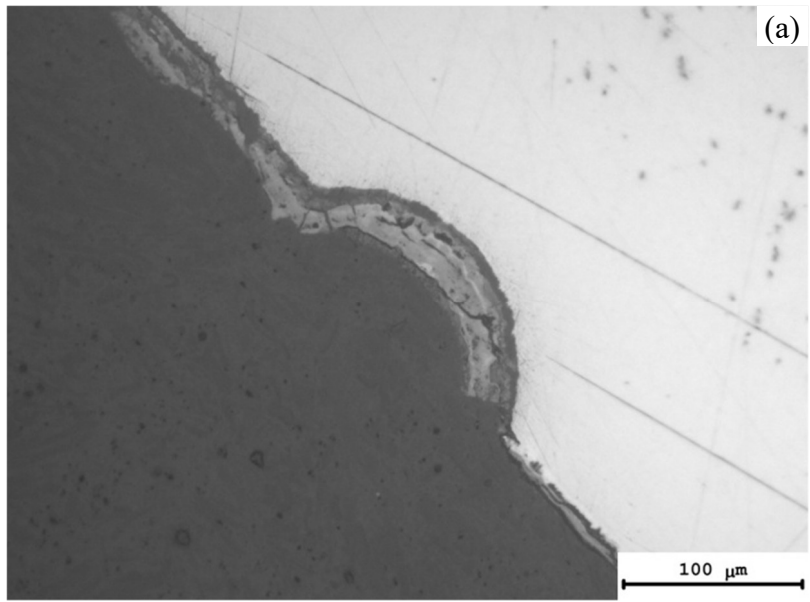
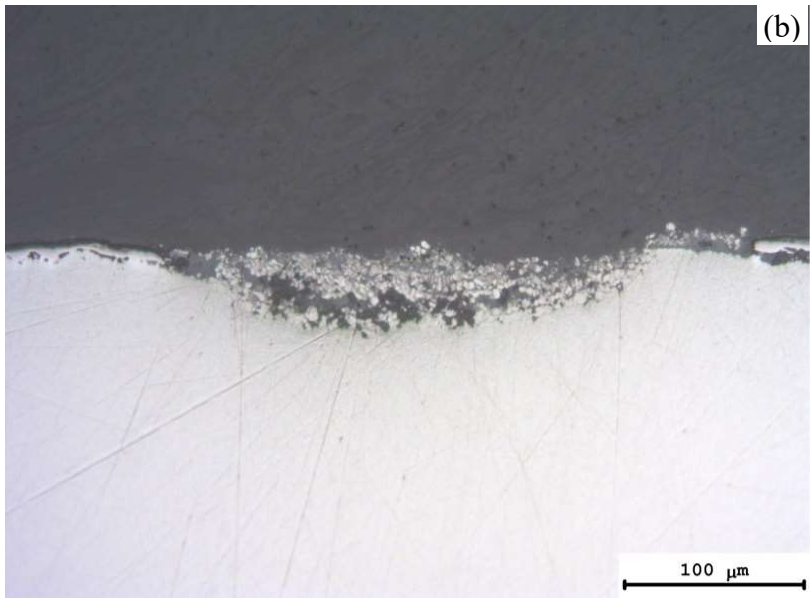
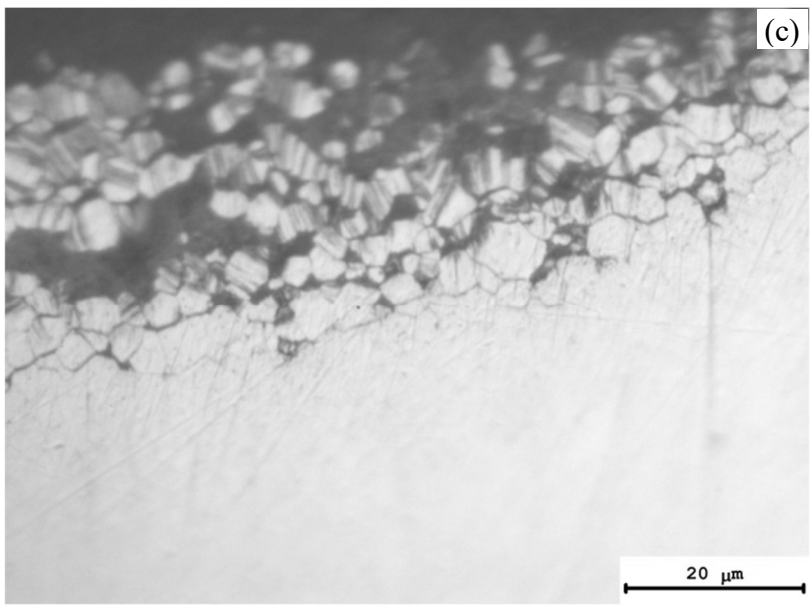

Figure 9

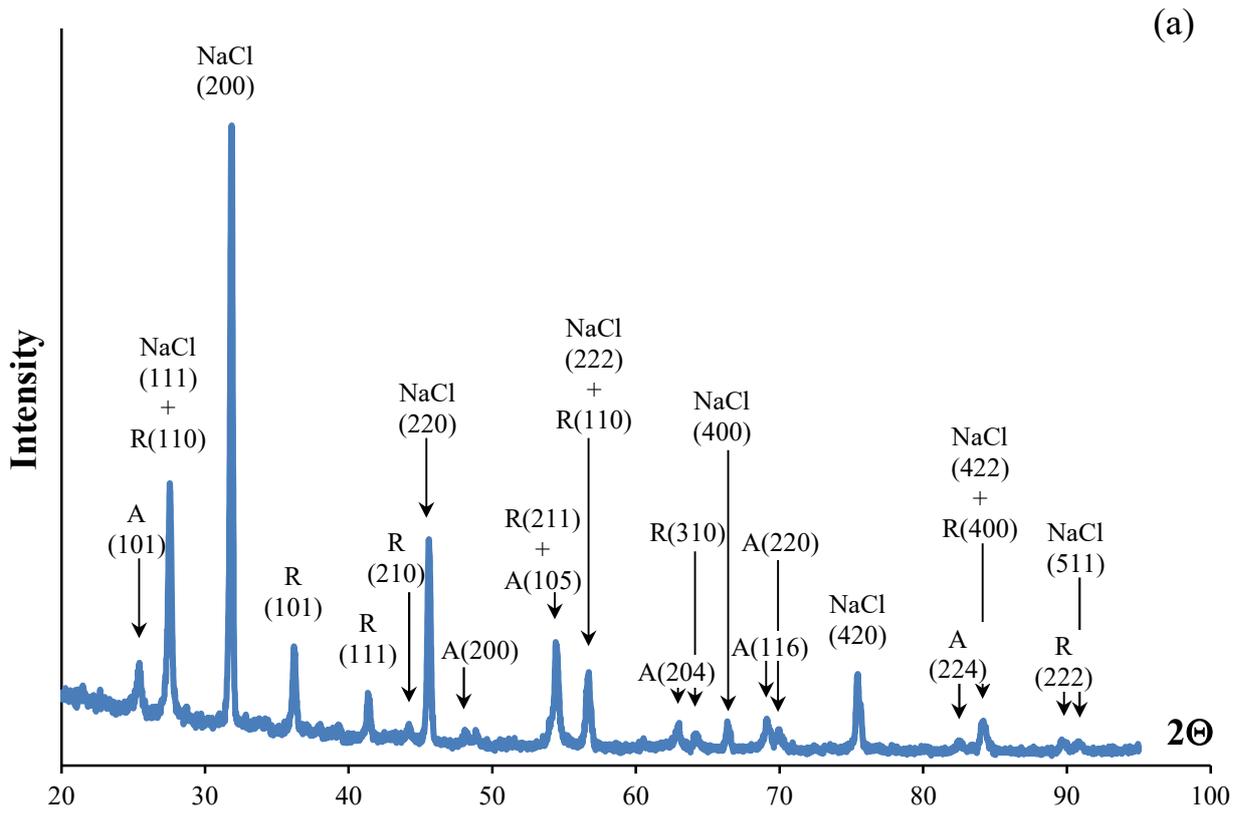

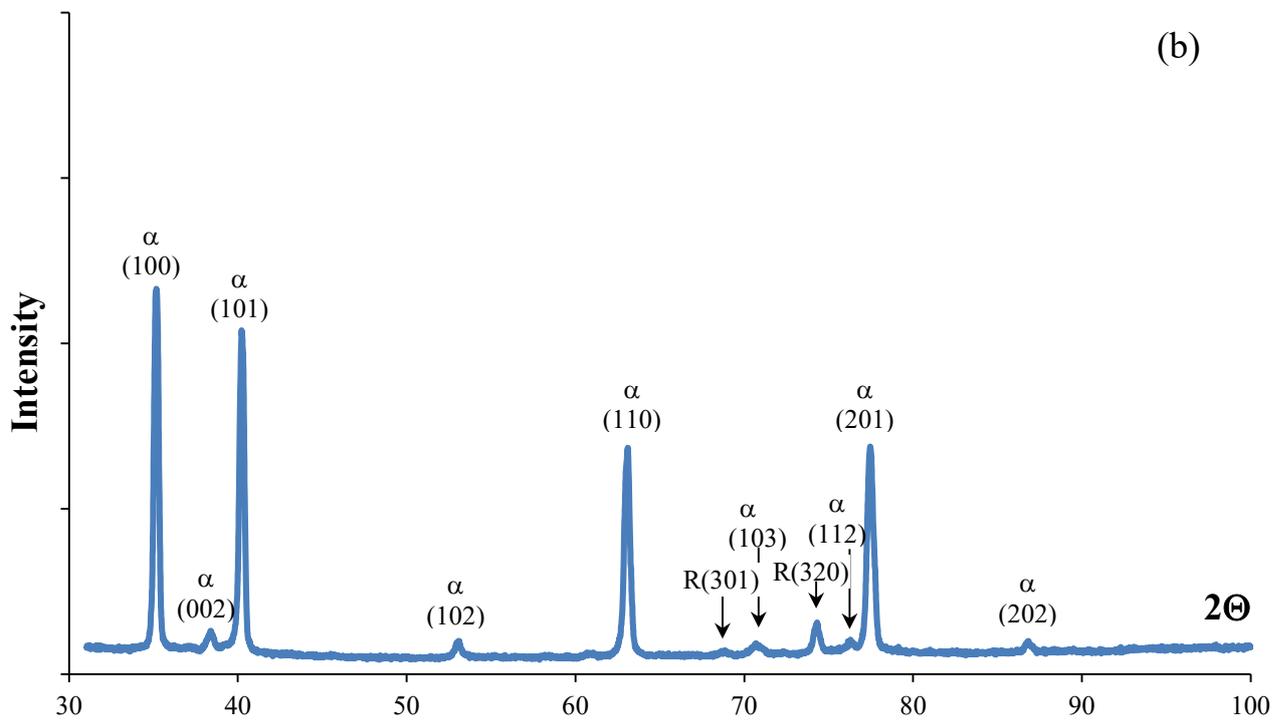

Figure 10

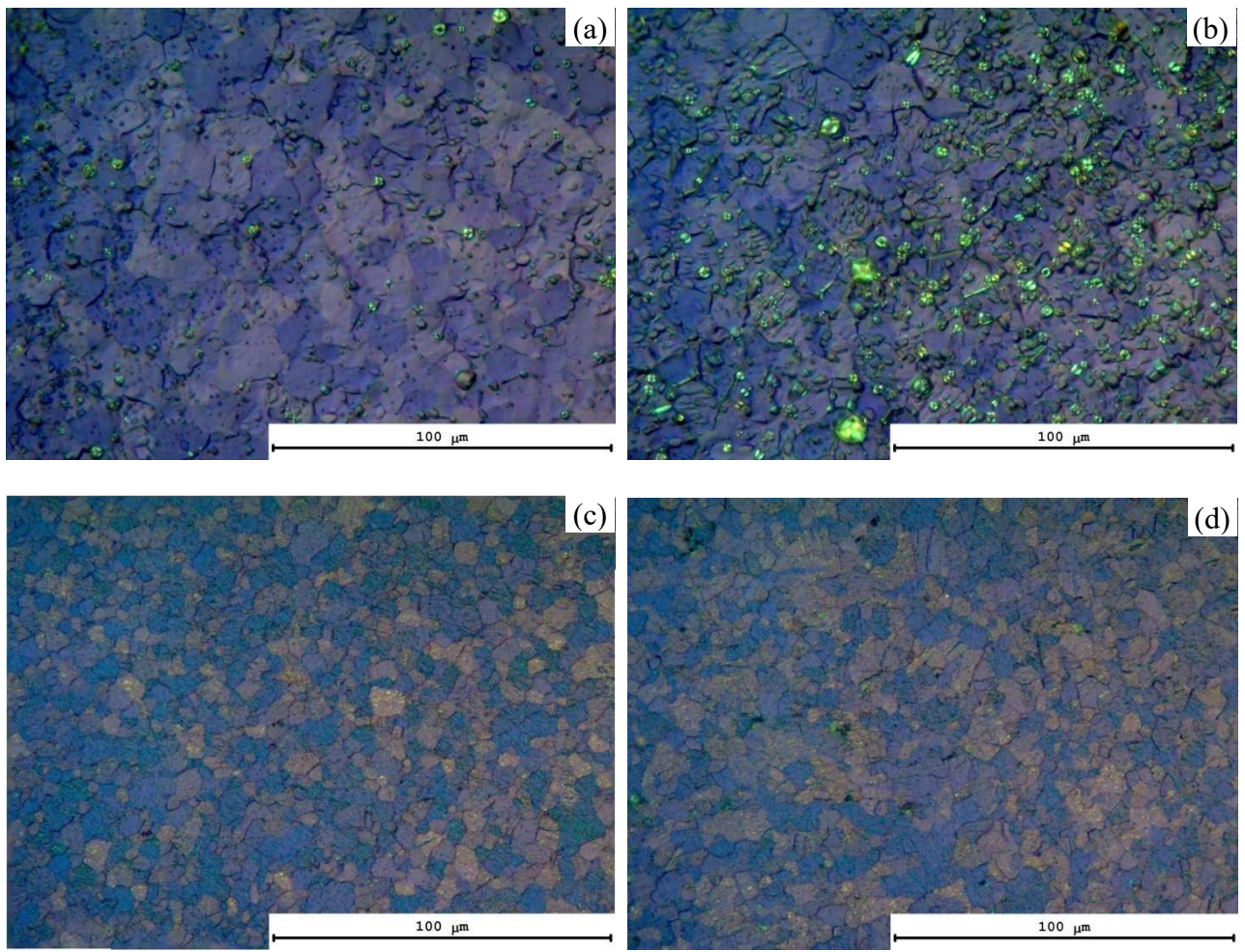

Figure 11

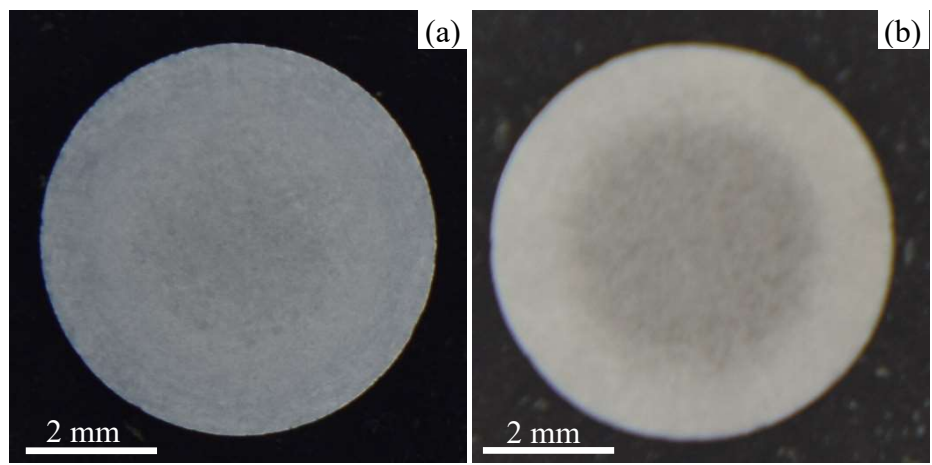

Figure 12

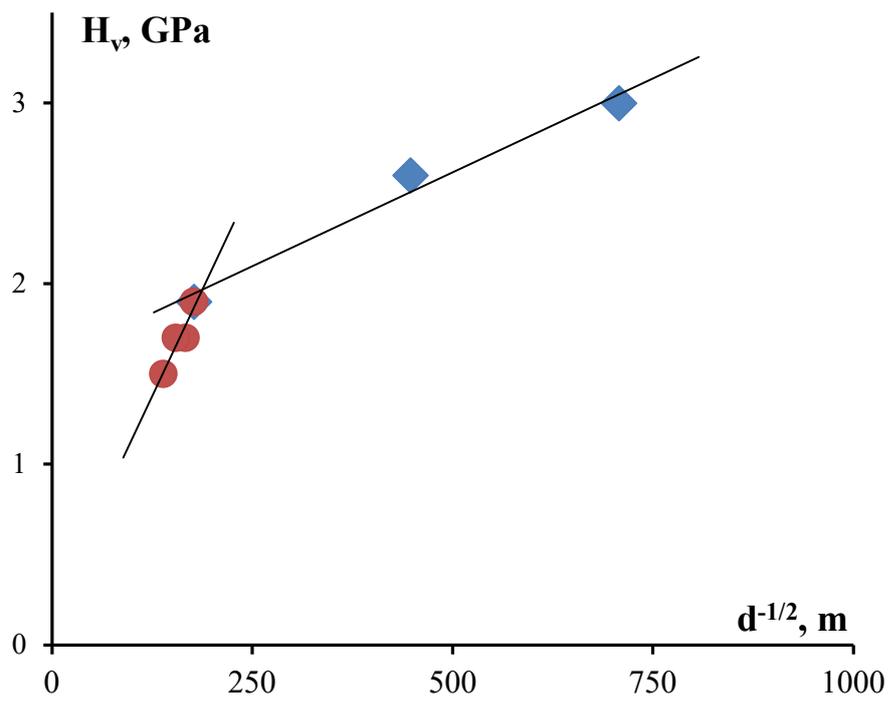

Figure 13

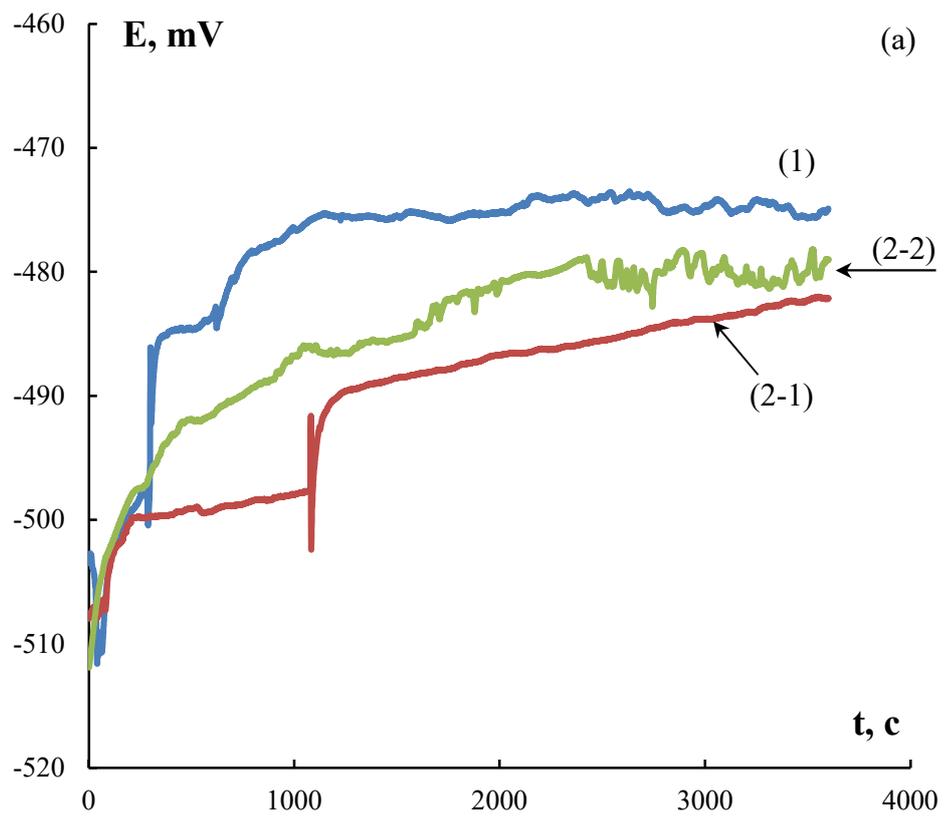
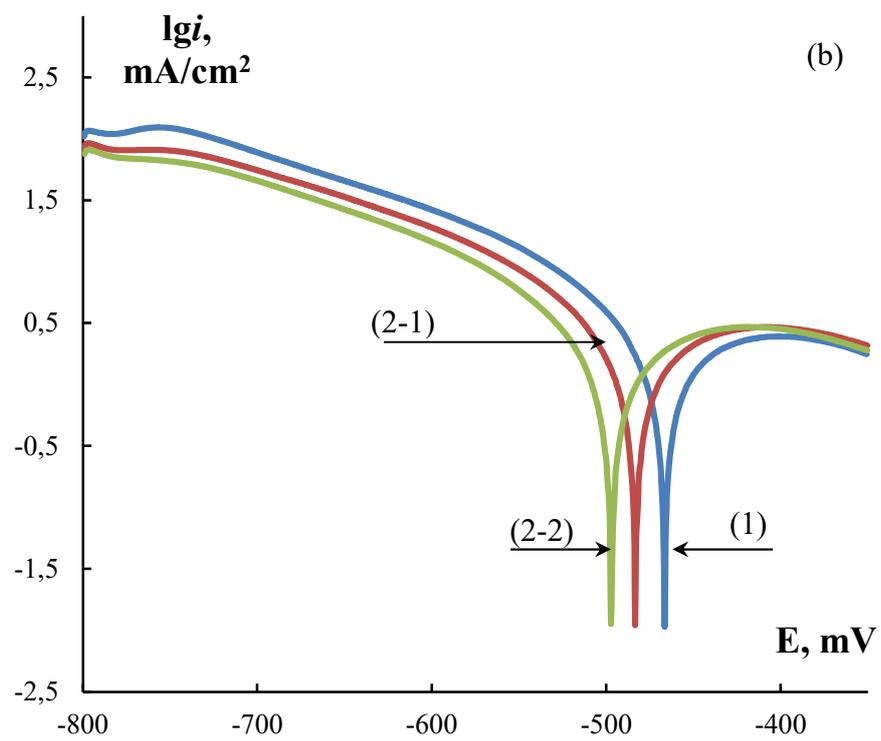

Figure 14

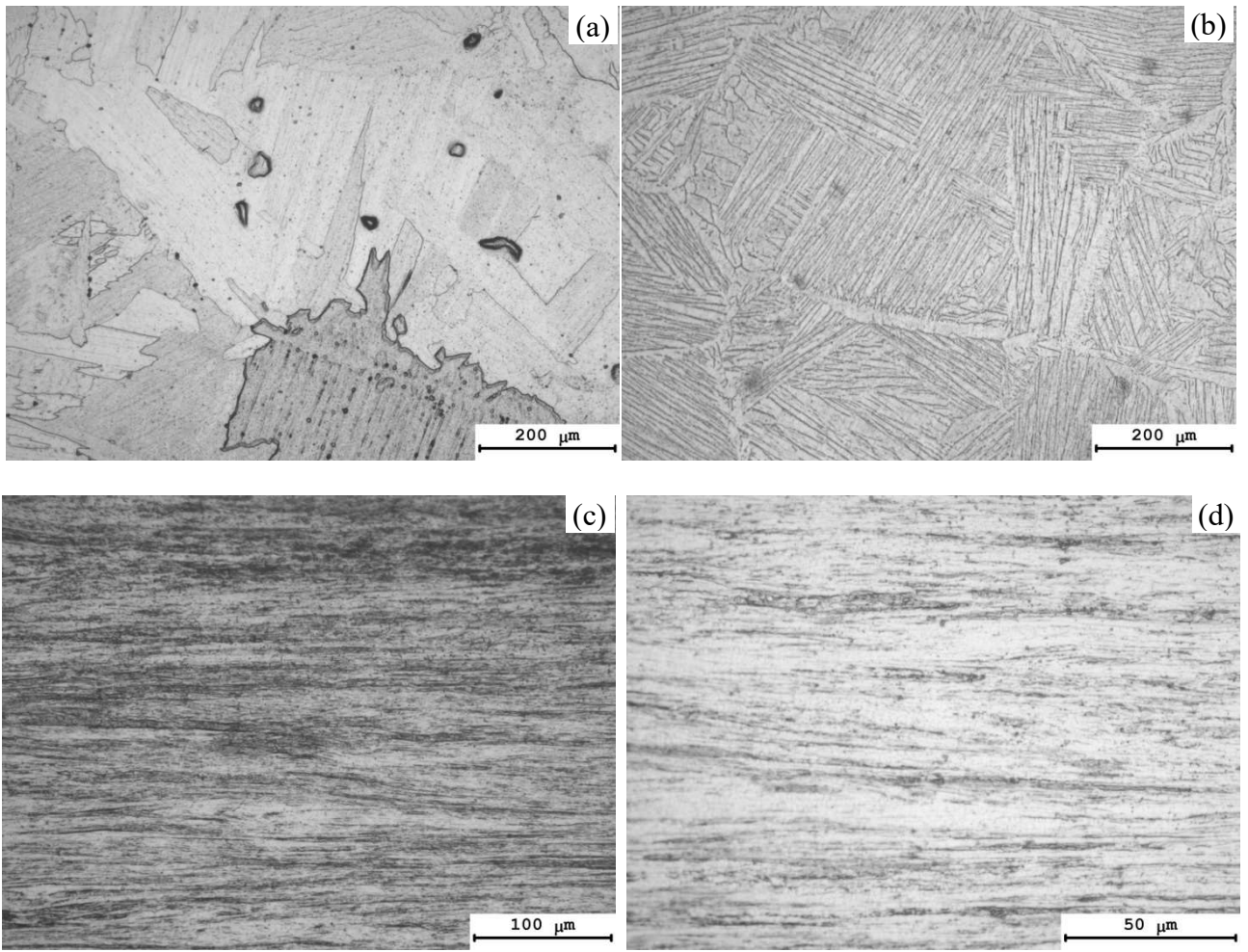

Figure 15